\documentclass[conference]{IEEEtran}
\IEEEoverridecommandlockouts
\usepackage{cite}
\usepackage{amsmath,amssymb,amsfonts}
\usepackage{algorithmic}
\usepackage{graphicx}
\usepackage{multirow}
\usepackage{nicematrix}
\usepackage{booktabs} 
\usepackage[normalem]{ulem} 
\usepackage{lipsum} % For dummy text, you can remove this in your actual document
\usepackage{subcaption}

\usepackage{textcomp}
\usepackage{xcolor}
\def\BibTeX{{\rm B\kern-.05em{\sc i\kern-.025em b}\kern-.08em
    T\kern-.1667em\lower.7ex\hbox{E}\kern-.125emX}}
\begin{document}

% ------
\title{Metric Learning with Progressive Self-Distillation for Audio-Visual Embedding Learning}

\author{
\IEEEauthorblockN{Donghuo Zeng*\thanks{* Corresponding author}} 
\IEEEauthorblockA{\textit{KDDI Research, Inc.} \\
Saitama, Japan \\
do-zeng@kddi-research.jp}
\and
\IEEEauthorblockN{Kazushi Ikeda} 
\IEEEauthorblockA{\textit{KDDI Research, Inc.} \\
Saitama, Japan \\
kz-ikeda@kddi-research.jp}
}

\maketitle

\begin{abstract}
Metric learning projects samples into an embedded space, where similarities and dissimilarities are quantified based on their learned representations. However, existing methods often rely on label-guided representation learning, where representations of different modalities, such as audio and visual data, are aligned based on annotated labels. This approach tends to underutilize latent complex features and potential relationships inherent in the distributions of audio and visual data that are not directly tied to the labels, resulting in suboptimal performance in audio-visual embedding learning.
To address this issue, we propose a novel architecture that integrates cross-modal triplet loss with progressive self-distillation. Our method enhances representation learning by leveraging inherent distributions and dynamically refining soft audio-visual alignments—probabilistic aligns between audio and visual data that capture the inherent relationships beyond explicit labels. 
Specifically, the model distills audio-visual distribution-based knowledge from annotated labels in a subset of each batch. This self-distilled knowledge is used to automatically generate soft-alignment labels for the remaining audio-visual samples. These soft-alignment labels are used to construct soft cross-modal triplets, which in turn are employed to fine-tune the model's parameters.
Experimental results on two audio-visual benchmark datasets demonstrate the effectiveness of our proposed method in the cross-modal retrieval task, achieving state-of-the-art performance with improvements of 2.13\% and 1.82\% on the AVE and VEGAS datasets, respectively, in terms of average MAP metrics.
% Supervised metric learning depends on accurate annotations to assess similarities and dissimilarities between samples. However, existing methods often fail to consider the potential semantic relationships among negative samples as noisy data, leading to suboptimal performance in audio-visual embedding learning. % Our method dynamically refines soft audio-visual alignments, enabling the model to learn robust shared representations from noisy data. Our approach distills its own audio-visual distribution-based knowledge from annotated labels in a subset of each batch to automatically generate soft-alignment labels for the remaining samples. These soft-alignment labels are used to construct soft cross-modal triplets,  which are then utilized to fine-tune the model's parameters for improved performance. 
\end{abstract}


\begin{figure}[!t]
\centering
\includegraphics[width=0.46\textwidth]{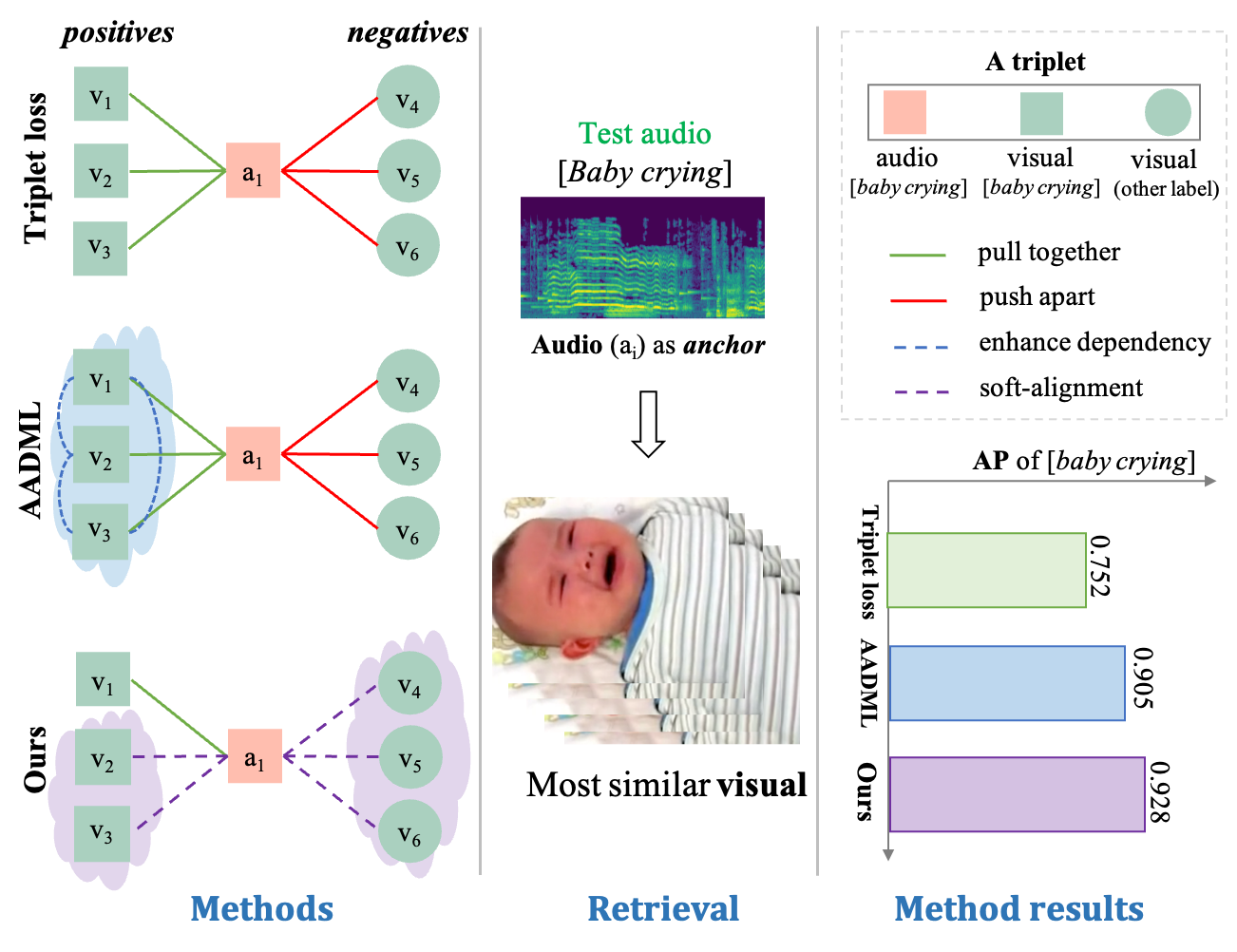}
\caption{The standard triplet loss pulls the positives closer to the anchor ($a_1$) and pushes the negatives away but struggles with limited data. AADML~\cite{zeng2024anchor} enhances dependencies between similar samples but overlooks the latent information inherent in the distributions beyond labels. Our approach addresses this by generating soft audio-visual alignments, enhancing embeddings for cross-modal tasks like retrieving \textbf{\textit{baby crying}} visuals from the audio on the VEGAS~\cite{zhou2018visual} and improving its average precision (AP).}
\label{fig:overview}
\vspace{-1.5pc}
\end{figure}

\section{Introduction}
% \textcolor{red}{\textbf{Note}: soft-alignment is the same as soft audio-visual alignment in this paper, I may change soft label to soft-alignment label. In summary, soft-alignment is the connection between audio and visual, which is probabilistic matched/mismatched. soft-alignment label is the predict label of probability.}
\label{sec:intro}
Metric learning~\cite{kulis2013metric, lowe1995similarity, weinberger2005distance, xing2002distance, hoffer2015deep, vinh2020hyperml, venkataramanan2021takes, li2022neighborhood, lee2008rank} is a powerful approach for learning representations by comparing sample similarities and dissimilarities based on annotated labels, such as triplet loss~\cite{hermans2017defense}. This technique projects samples into an embedded space, where the relationships between samples are quantified according to their learned representations, which can significantly improve tasks such as audio-visual embedding for cross-modal retrieval~\cite{zeng2019learning, zeng2020deep, zeng2023learning}.
% . A widely used metric learning technique is triplet loss~\cite{hermans2017defense}, which integrates the optimization of positive and negative samples within a single loss function. 
% Specifically, triplet loss directly optimizes the model by pulling positive samples closer together while pushing negative samples further apart in one formulation. 
% This direct optimization process helps in learning a discriminative feature space that can significantly improve tasks such as audio-visual embedding for cross-modal retrieval~\cite{zeng2020deep, zeng2023learning}.
% However, metric learning faces a significant challenge in practical scenarios: its heavy reliance on annotated labels. 
However, metric learning often has a heavy reliance on annotations, which limits the effectiveness and scalability of the approach, as obtaining high-quality annotations is often labor-intensive and costly. Furthermore, the simplicity of these annotations means that label-guided representation learning often cannot achieve acceptable performance on its own, as it may not capture the complex, latent features of the data.% Since positive and negative samples are derived from annotations, the imbalance—where negative samples often vastly outnumber positive ones—can lead to suboptimal embedding spaces and difficulties in model convergence, ultimately impeding performance.

Existing methods~\cite{zheng2019hardness, zeng2022complete} attempt to mitigate these issues by focusing on selecting impactful data samples for the representation learning on embedding space. However, these approaches fail to fully explore the space due to the limited training data, resulting in an incomplete representation of samples. The state-of-the-art (SOTA) model, AADML~\cite{zeng2024anchor}, addresses this issue by discovering the potential correlations among similar samples to enhance the quality of representations (see Fig.~\ref{fig:overview}).
% , thereby enhancing the quality of the representation learning on shared embedding space for audio-visual learning. 
Nonetheless, AADML does not effectively take into account the valuable latent features and potential relationships inherent in the data distributions. Specifically, when determining positive and negative samples using fixed annotations, this approach treats all positives and negatives with equal importance. Consequently, it neglects the distributional nuances of the data, which are essential for more accurately identifying and differentiating between positive and negative samples.

To address this issue, we propose a novel architecture that integrates cross-modal triplet loss with progressive self-distillation. The goal is to iteratively refine the model's knowledge to soften rigid annotations and capture nuanced relationships in the data. Our method improves the learning process by efficiently utilizing inherent data distribution and dynamically refining soft audio-visual alignments—probabilistic associations that capture relationships beyond explicit label constraints. 
% probabilistic matches between audio and visual data that capture the inherent relationships beyond explicit labels. 
Each batch is partitioned into two subsets: the first is used to project audio-visual data into a common feature space by minimizing the distance between learned features and ground truth labels. The model then generates soft-alignment labels for the second subset, which are used to create soft cross-modal triplets by identifying positive and negative samples for loss optimization. To preserve the original pairs, we reduce cross-modal dependency by minimizing the distance between paired audio and visual representations. As training progresses, the partitioning rate decreases, allowing the teacher model to generate soft-alignment labels for progressively larger portions of each batch, gradually transforming the model into its own teacher. 
% This iterative process enhances learning from noisy data and improves overall performance. Specifically, each batch is randomly partitioned into two subsets. The model projects audio-visual data of the first subset into a common feature space by minimizing the distance between learned features and ground truth labels. This model is used to generate soft-alignment labels for the second subset. These soft-alignment labels form soft cross-modal triplets, with positives and negatives derived from them, and are utilized for loss optimization. Additionally, to reduce cross-modal dependency, we minimize the distance between paired audio and visual representations. This approach enables our method to continuously model new soft cross-modal triplets while recalibrating potentially poorly matched or mismatched samples without explicit identification. As training progresses, the partitioning rate decreases, allowing the teacher model to generate soft-alignment labels for progressively larger portions of each batch, enabling the model network to transition into its own teacher. This iterative process enhances learning from noisy data and improves overall model performance.

To validate the effectiveness of our approach, we conduct experiments on two datasets.
% , each containing shared annotation labels for both audio and visual modalities. 
The results demonstrate the superior performance of our method, with significant improvements over SOTA. Specifically, our approach achieves a 2.13\% increase in mean average precision (MAP) on the AVE~\cite{tian2018audio} dataset and a 1.82\% increase on the VEGAS~\cite{zhou2018visual} dataset in audio-visual cross-modal retrieval (AV-CMR).
% In this paper, we identify and address this gap by proposing a novel architecture that integrates cross-modal triplet loss with progressive self-distillation. Our method not only refines the handling of positive samples but also dynamically addresses the noisy nature of negative samples. By leveraging soft audio-visual alignments and iteratively updating the model, our approach aims to capture and utilize the potential information from the negative samples more effectively, thereby enhancing audio-visual embedding learning beyond the capabilities of existing methods.

\section{Related Work}
\label{sec:rela}

\subsection{Audio-visual Embedding Learning} Audio-visual embedding learning creates a shared feature space for audio and visual data, preserving semantic relationships and enabling effective cross-modal retrieval~\cite{zeng2022complete}. CCA\cite{hardoon2004canonical} and its variants, including K-CCA\cite{akaho2006kernel}, Cluster-CCA~\cite{rasiwasia2014cluster}, DCCA~\cite{andrew2013deep} and C-DCCA~\cite{yu2018category}.
% , are commonly used to learn joint representations that capture correlations between ？modalities. 
These methods often struggle to capture complex non-linear relationships.
% However, these methods often struggle with capturing intricate, non-linear relationships present in real-world data. 
Deep learning excels in audio-visual embedding by aligning matched samples using neural networks.
% Deep learning has become a dominant approach for audio-visual embedding learning, employing neural networks to construct joint embeddings that reduce the distance between matched samples. 
% Models like DSCMR~\cite{zhen2019deep}, 
Models like  CCTL~\cite{zeng2022complete}, and EICS~\cite{zeng2023learning} capture intricate audio-visual relationships to advance retrieval tasks. 
% While previous work has focused on sophisticated neural networks for audio-visual representation learning, our approach, inspired by AADML~\cite{zeng2024anchor}, 
Inspired by AADML~\cite{zeng2024anchor}, we introduce a novel metric learning that leverages latent features and potential relationships beyond annotation through self-knowledge distillation.

\subsection{Self-knowledge Distillation} 
Knowledge distillation (KD)~\cite{hinton2015distilling} typically involves a large teacher model guiding a smaller student model. In cross-modal tasks, KD~\cite{gou2021knowledge} transfers knowledge from a pre-trained model on one modality to a student model with a new modality, often using paired samples. 
% Various methods~\cite{gou2021knowledge} utilize paired samples, contrastive loss, and modality hallucination to improve performance across different tasks, such as pose estimation and action recognition. 
Various methods~\cite{gou2021knowledge} enhance performance using paired samples, contrastive loss, and modality hallucination.
Despite its success in visual recognition, cross-modal knowledge transfer struggles with modality gaps and paired sample availability. We introduce progressive self-distillation to iteratively refine the model’s knowledge, and soften constraints to capture transferable cross-modal representations.
% Our approach is inspired by recent advancements~\cite{hahn2019self, gou2021knowledge} in self-knowledge distillation, where the student network acts as its own teacher, we introduce progressive self-distillation for audio-visual embedding learning.

\begin{figure*}[t]
\centering
\includegraphics[width=0.9\textwidth]{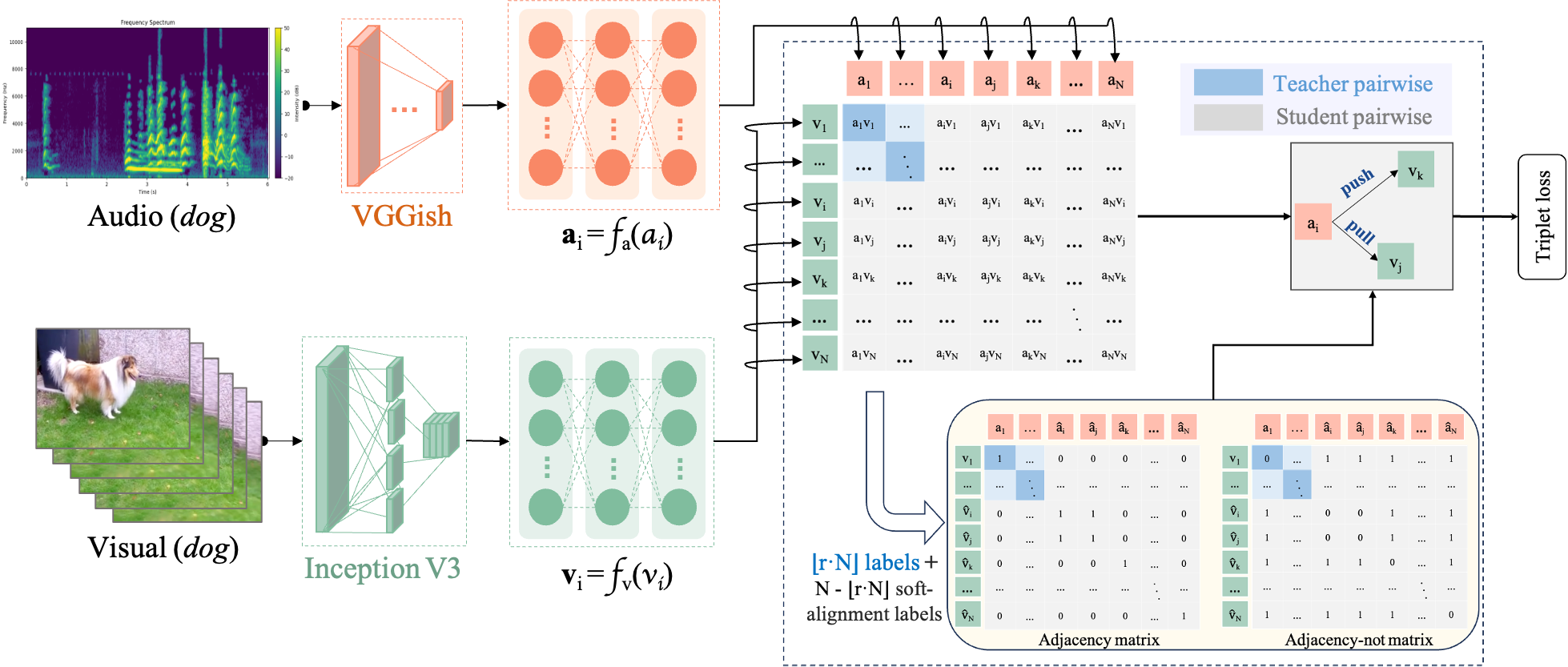}
\caption{The overview of our approach. The batch is divided into annotated instances (light blue background) and unannotated instances (light grey background) in the above three matrices by the hyper-parameter $r$. The teacher network is trained on annotated instances using the cross-modal triplet loss. The teacher estimates soft-alignment labels for the unannotated data, which are then used to supervise the student network. As training progresses and the teacher's representations become more reliable, the proportion of soft-alignment labels provided to the student increases.}
\label{fig:architecture}
\vspace{-1.2pc}
\end{figure*}

\section{Approach}
\label{sec:app}
To address the issue of overlooking inherent distribution beyond annotations, we propose a model that progressively distills its own knowledge to generate soft-alignment labels and enabling more transferable representations for cross-modal triplet learning, as seen in Fig.~\ref{fig:architecture}.
\vspace{-0.2pc}
\subsection{Preliminaries}
% Audio-visual cross-modal embedding learning presents challenges due to the inherently distinct and complex feature distributions of audio and visual data, which cannot be directly compared using standard metrics. To address this, it is necessary to project the audio-visual data into a common feature space to enable feasible comparisons.
% Audio-visual embedding learning is challenging due to the distinct and complex feature distributions of audio and visual data, which cannot be directly compared. Projecting the data into a common feature space is necessary for feasible comparisons. 
Consider an audio-visual dataset comprising $n$ videos, denoted as $\mathcal{D} = \{ (a_{i}, v_{i}) \}_{i=0}^{n-1}$, where each video pair $(a_{i}, v_{i})$ is represented as ($a_{i}$, $v_{i}$). The $a_{i} \in R^{128}$ denotes the audio feature extracted using a VGGish model, and $v_{i} \in R^{1024}$ denotes the visual features extracted using an Inception V3 model. Each pair is associated with a binary semantic label vector $Y_{i} \in \{0, 1\}^{c}$, where $c$ is the number of categories, 1 indicates that the video pair $(a_{i}, v_{i})$ belongs to the corresponding class, while 0 indicates otherwise. The goal of our model is to train encoders $f_a$ for audio data and $f_v$ for visual data such that for a given semantically similar instance pairs $(f_a(a_i), f_v(v_j))$ with the same label get closer together while dissimilar instance pairs with different labels push farther apart.

% We project part of the \(a_i\) and \(v_i\) into a common feature space aligned with predefined labels to train the model, then used to predict soft-alignment labels for the rest samples for building cross-modal triplet for the metric learning optimization.

% We aim to project $a_{i}$ and $v_{i}$ into a unified embedding space that aligns with predefined labels. Our approach enhances the model's ability to create accurate and comprehensive semantic representations, significantly improving metric learning performance by considering the potential semantic relationships between negative samples.

\subsection{Soft Cross-modal Triplet}
A cross-modal triplet (e.g., \((a_i, v_i^+, v_i^-)\)) is composed of a sample (audio $a_i$) as the anchor from one modality and two samples as positive (visual $v_i^+$) and negative (visual $v_i^-$) from another modality. The purpose of a cross-modal triplet is to ensure that the positive visual sample \(v_i^+\) is closer to the anchor \(a_i\) than the negative visual sample \(v_i^-\) within a common feature space, thereby improving the model’s ability to differentiate relevant from irrelevant content across modalities. In relatively large batch settings, relying solely on annotated labels to determine positive and negative samples will overlook the varying degrees of similarity and dissimilarity between the audio and visual samples. This is because such an approach fails to capture potential relationships beyond the labels, making it difficult to accurately define $v_i^+$ and $v_i^-$. 
% Additionally, incorrectly or loosely correlated pairs may lead to underutilizing the complexity of the relationships, limiting the accuracy of triplet information.
% Consequently, building cross-modal triplets based solely on fixed annotated labels may fail to reflect the true degrees of relatedness between samples and overlook important content not directly tied to the annotated labels.
% To address this issue, our approach generates soft-alignment labels for the audio-visual data by leveraging the inherent distributions, allowing a well-trained teacher model to refine poorly matched visual samples by re-pairing them with stronger semantic matches or mismatches within the batch. This process dynamically redefines $v_i^+$ and $v_i^-$ to better capture potential content beyond the annotated labels, thereby reducing the reliance on these annotated labels.
To address this issue, our approach generates soft-alignment labels for audio-visual data by capturing latent features and potential relationships beyond explicit annotations. This process allows the teacher model to iteratively refine the classification of samples by adjusting the identification of soft positives $\hat{v}_i^+$ and soft negatives $\hat{v}_i^-$ based on the learned distributions. The soft-alignment labels guide the model in distinguishing subtle variations in semantic relationships, leading to improved matching and overall performance.

Soft-alignment labels are generated by the teacher model trained on a subset of every batch. The teacher model consists of audio and visual encoders $f_{a}(\cdot)$ and $f_{v}(\cdot)$, the output logits $\mathbf{\hat{A}} = \{\hat{a}_i\} \in \mathbb{R}^{N \times 128}$ and $\mathbf{\hat{V}} = \{\hat{v}_i\} \in \mathbb{R}^{N \times 1024}$ used to produce soft-alignment label distributions \textbf{$\mathbf{L^{a}}$} and \textbf{$\mathbf{L^{v}}$}, which defined as:
\begin{equation}
\begin{aligned}
\mathbf{L^{a}} = \sigma (\hat{V}\cdot \hat{A}^{T}) \text{ and } 
\mathbf{L^{v}} = \sigma (\hat{A}\cdot \hat{V}^{T})
\end{aligned}
\end{equation}
where $\sigma(\cdot)$ is the softmax function to transform raw logits into probability. 
% \(\{ \mathbf{z}_i^a \}\) and \( \{\mathbf{z}_i^v \}\) used to produce probability distributions \textbf{$\mathbf{\hat{a}}_{i}$} and \textbf{$\mathbf{\hat{v}}_{i}$}.
% $\mathbf{P^a}$=$\{p_i^a\}$ and $\mathbf{P^v}$=$\{p_i^v\}$, which indicate the likelihood of each sample belonging to each possible label. These distributions reflect the degree of alignment between \(a_i\) and each \(v_i\), allowing for a more flexible and nuanced definition of \(v^+\) and \(v^-\). The soft-alignment labels help the model capture underlying semantic relationships that may not be explicitly labeled, improving the quality of the cross-modal triplets. The audio probability distributions are defined as:
% \begin{equation}
% \begin{aligned}
% p_i^a(k) = \frac{e^{z_i^a(k)}}{\sum_{l=1}^C e^{z_i^a(l)}}
% \end{aligned}
% \end{equation}
% where $\mathbf{p}_i^a = \left[p_i^a(1), p_i^a(2), \ldots, p_i^a(C)\right]$,  \( z_i^a(k) \) is the logit for class \( k \) of the audio sample \( a_i \), and \( C \) is the total number of categories. Similarly, the visual probability distribution \( \mathbf{p}_i^v \) is computed using the same softmax function applied to the logits \( \mathbf{z}_i^v \) for visual samples. 
The soft label distributions are used to construct soft cross-modal triplets by incorporating the adjacency matrix \(A\) and the adjacency-not matrix \(\bar{A}\) into the loss calculation. They are defined as:
\begin{equation}
\begin{aligned}
  A_{ij} =
\begin{cases}
1 & \text{if } \text{argmax}(\mathbf{L}_j^a) = \text{argmax}(\mathbf{L}_i^v) \\
0 & \text{otherwise}
\end{cases} \\
\bar{A}_{ij} =
    \begin{cases}
    1 & \text{if } \text{argmax}(\mathbf{L}_j^a) \neq \text{argmax}(\mathbf{L}_i^v) \\
    0 & \text{otherwise}
    \end{cases} 
\end{aligned}
\end{equation}
where the $L^a_i$ refers to the i-th column of $L^a$ and $L^v_j$ corresponds to the j-th row of $L^v$. Incorporate these matrices into the triplet loss to identify the positive visual $v_i^+ =\hat{v}_j$ ($A_{ij}=1$) and the negative visual $v_i^- =\hat{v}_k$ ($\bar{A}_{ik}=1$) for $a_i$. By using the $AA(\cdot)$ proxy~\cite{zeng2024anchor} for each representation to enhance the correlation between positive samples, our final cross-modal triplet loss can be formulated as:
\begin{equation}
\begin{aligned}
l_{cross}=\sum_{i=0}^{N-1}\max\{0, \, d(AA(a_i), AA(v_i^+)) \\
- d(AA(a_i), AA(v_i^-)) + \alpha\}
\end{aligned}
\end{equation}
where $d(\cdot)$ are the normalized Euclidean distance, $N$ is the batch size.

\subsection{Progress Self-distillation}
Self-knowledge distillation~\cite{furlanello2018born,hahn2019self, gou2021knowledge} allows a student network to act as its own teacher, cutting computational costs compared to traditional methods. Progressive self-knowledge distillation enhances this by gradually evolving the student into its own teacher during training, leading to a more dynamic and efficient learning process.
% Self-knowledge distillation enables a student network to serve as its own teacher, thereby reducing computational costs compared to traditional knowledge distillation methods. Progressive self-knowledge distillation further optimizes this approach by incrementally evolving the student network into its own teacher throughout the training process. 
% Unlike conventional methods where the teacher model remains static and distinct from the student, this adaptive strategy fosters a more dynamic and efficient learning experience. 
Our progressive self-distillation process begins with a batch of $N$ samples, which is randomly partitioned into two subsets: $N_{1} = \lfloor \text{r} \cdot N \rfloor$ samples with ground-truth labels and $N_{2} = N-\lfloor \text{r} \cdot N \rfloor$ samples without labels. Initially, the subset $N_{1}$ is used to train the teacher network with ground truth pairings. The teacher network then generates soft-alignment labels for the subset $N_{2}$, which are used to supervise its own student model. 

To enhance the teacher's influence on learning, we progressively reduce $r$. This gradual decrease ensures that the student network increasingly relies on its own predictions over time. We use a step-wise schedule to adjust $r$ from a specified start value (1.0) to an end value (0.2), as this method has proven more effective for our model than other strategies, such as linear or cosine-annealing schedule~\cite{li2020few}.
% Various strategies can be employed for decreasing $r$, including step-wise, linear, or cosine-annealing schedules. In our approach, use a step-wise schedule, adjusting $r$ from a specified start value to an end value, as this method proves most effective for our model.

By incorporating this progressive self-distillation approach, our model dynamically evolves and improves its representational capabilities. In addition, we reduce the cross-modal dependency of exact audio-visual pairs while preserving the original alignment, achieved by $l_{dis}$ loss.
% , resulting in enhanced performance and efficiency throughout the learning process. 
The final objective loss is defined as follows:
\begin{equation}
\begin{split} 
   Loss &= l_{lab} + l_{cross} + l_{dis}\\
   l_{lab} &= \frac{1}{n}||f_{a}(a_{i})-Y(a_{i})||_{F} + \frac{1}{n}||f_{v}(v_{i})-Y(v_{i})||_{F} \\
   l_{dis} &= \frac{1}{n}||f_{a}(a_{i})-f_{v}(v_{i})||_{F}
    \label{eq:loss}
\end{split}
\vspace{-1.5pc}
\end{equation}
where $||\cdot||_{F}$ signifies the Frobenius norm, $f(x)$ represents the projected feature in the shared label space and $Y(\cdot)$ denotes the label representations. The final objective loss is optimized using the stochastic gradient descent (SGD)~\cite{ruder2016overview}.

\section{Experiments}
\label{sec:expe}

\begin{table}[t]
\centering
\small % Adjusts the font size for the table
\caption{Comparison of MAP between our approach and existing methods (The best in bold and second-best underlined)}
\begin{NiceTabular}{c@{\hspace{0.05em}}|@{\hspace{0.1em}}ccc@{\hspace{0.1em}}|@{\hspace{0.1em}}ccc}
\toprule
\Block{2-1}{\textbf{Models}} &  \Block{1-3}{\textbf{AVE Dataset}} &&  &\Block{1-3}{\textbf{VEGAS Dataset}} \\ \cline{2-7}
                       & A$\rightarrow$V & V$\rightarrow$A  & Avg.
                       & A$\rightarrow$V & V$\rightarrow$A  & Avg. \\ \hline
    Random
    & 0.127 & 0.124 & 0.126 & 0.110 & 0.109 & 0.109     \\
    CCA~\cite{hardoon2004canonical} 
    & 0.190 & 0.189 & 0.190 & 0.332 & 0.327  & 0.330    \\
    DCCA~\cite{andrew2013deep}  
    & 0.221 & 0.223 & 0.222 & 0.478 & 0.457  & 0.468  \\
    C-CCA~\cite{rasiwasia2014cluster}  
    & 0.228 & 0.226 & 0.227 & 0.711 & 0.704  & 0.708    \\
    % C-DCCA~\cite{yu2018category}
    % & 0.230 & 0.227 & 0.229 % & 0.722 & 0.716  & 0.719    \\
    TNN-CCCA~\cite{zeng2020deep} 
    & 0.253 & 0.258 & 0.256 & 0.751 & 0.738 & 0.745      \\ 
    % DSCMR~\cite{zhen2019deep}  
    % & 0.314 & 0.256 & 0.285& 0.732 & 0.721 & 0.727      \\
    CCTL~\cite{zeng2022complete}   & 0.328 & 0.267 & 0.298 & 0.766 & 0.765 & 0.766 \\ 
    V-Adviser~\cite{wang2023videoadviser}  &- &- &- &0.825 & 0.819 & 0.822 \\
    EICS~\cite{zeng2023learning}   &0.337 & 0.279 & 0.308 & 0.797 & 0.779 & 0.788\\
    TLCA~\cite{zeng2023two}  &0.410 &0.451 &0.431 &0.822 &0.838 &0.830\\
    % MSNSCA~\cite{zhang2023multi}  &0.323 &0.343 &0.333 &0.866 &0.865 &0.866\\
    AADML~\cite{zeng2024anchor}  &\uline{0.890} &\uline{0.883} &\uline{0.887} &\uline{0.901} &\uline{0.891} &\uline{0.896}\\ \hline
    \textbf{Ours}  &\textbf{0.909} &\textbf{0.907} &\textbf{0.908} &\textbf{0.918} &\textbf{0.910} &\textbf{0.914} \\
    \bottomrule
\end{NiceTabular}
\label{tab:comparison}
\vspace{-1.0pc}
\end{table}

\subsection{Datasets and Metrics}
Our model effectively performs the AV-CMR task by leveraging the assumption that audio and visual modalities share identical semantic information. We use two audio-visual datasets, VEGAS~\cite{zhou2018visual} and AVE~\cite{tian2018audio}, where the VEGAS includes 28,103 videos with 10 labels, and the AVE comprises 1,955 videos with 15 labels. We follow the same data partitioning and feature extraction methods as detailed in prior work~\cite{zeng2020deep}. We adopt the mean average precision (MAP) used in related works~\cite{zeng2023learning, zeng2022complete} as the evaluation metric.

\subsection{Implementation Settings}
Our model incorporates three fully connected (FC) layers and a prediction layer for audio and visual inputs Each FC layer has 1024 hidden units with a dropout rate of 0.1 to prevent overfitting. The dimensionality of the projected features matches the number of labels: 15 for AVE and 10 for VEGAS, aligning audio and visual features in a common space.
% In our model, we incorporate three fully connected (FC) layers and a prediction layer as decoders for both audio and visual inputs. Each FC layer is designed with 1024 hidden units, and we apply a dropout rate of 0.1 to mitigate overfitting. To align the audio and visual features within a common space, we ensure that the dimensionality of the projected features corresponds to the number of labels: 10 for the VEGAS dataset and 15 for the AVE dataset. 
The model is trained with batch sizes of 400 for both VEGAS and AVE, over 1,000 epochs. The triplet loss function is optimized using margin values of 1.2. We utilized the Adam optimizer~\cite{kingma2014adam} with default parameters, setting the learning rate at 0.0001 for all training procedures. These settings are consistent with the baselines in Table~\ref{tab:comparison}. Our experiments were conducted using the PyTorch and run on an Ubuntu Linux 22.04.2 system with an NVIDIA GeForce 3080 (10 GB) GPU. 
\subsection{Results}
To evaluate our method, we compared it with 9 algorithms: CCA-variants and deep learning (DL)-based methods. CCA-variants includes CCA~\cite{hardoon2004canonical}, and TNN-C-CCA~\cite{zeng2020deep}. DL-based methods contain
% DSCMR~\cite{zhen2019deep}, 
EICS~\cite{zeng2023learning}, VideoadViser~\cite{wang2023videoadviser}, 
% MSNSCA~\cite{zhang2023multi}, 
CCTL~\cite{zeng2022complete}, TLCA~\cite{zeng2023two}, and AADML~\cite{zeng2024anchor}.
We applied our approach to the AV-CMR task and compared it against the aforementioned methods on two audio-visual datasets shown in Table~\ref{tab:comparison}. Our method outperforms the others in all MAPs on both datasets, achieving gains of 
1.9\%, 2.4\%, and 2.1\% on AVE, and 1.7\%, 1.9\%, and 1.8\% on VEGAS in terms of A2V, V2A, and Average, respectively. These results demonstrate the effectiveness of our proposed model.

% \begin{table}[ht]
% \centering
% \begin{tabular}{|l|c|}
% \hline
% \textbf{Hyperparameter Name} & \textbf{Value / Function} \\
% \hline
% Batch size    & 400 \\
% Epoch         & 1,000 \\
% Learning rate & 0.0001 \\
% Audio NN      & [1024, 1024, 1024] \\
% Visual NN     & [1024, 1024, 1024] \\
% Dropout (NNs) & 0.1 \\
% Activate (NNs) & tanh \\
% Heads in Attention & 32 \\
% Margin & 1.2 \\
% \hline
% \end{tabular}
% \caption{Hyper-parameters}
% \label{table:parameter}
% \end{table}
\begin{figure}[t]
    \centering
    \includegraphics[width=1.0\columnwidth]{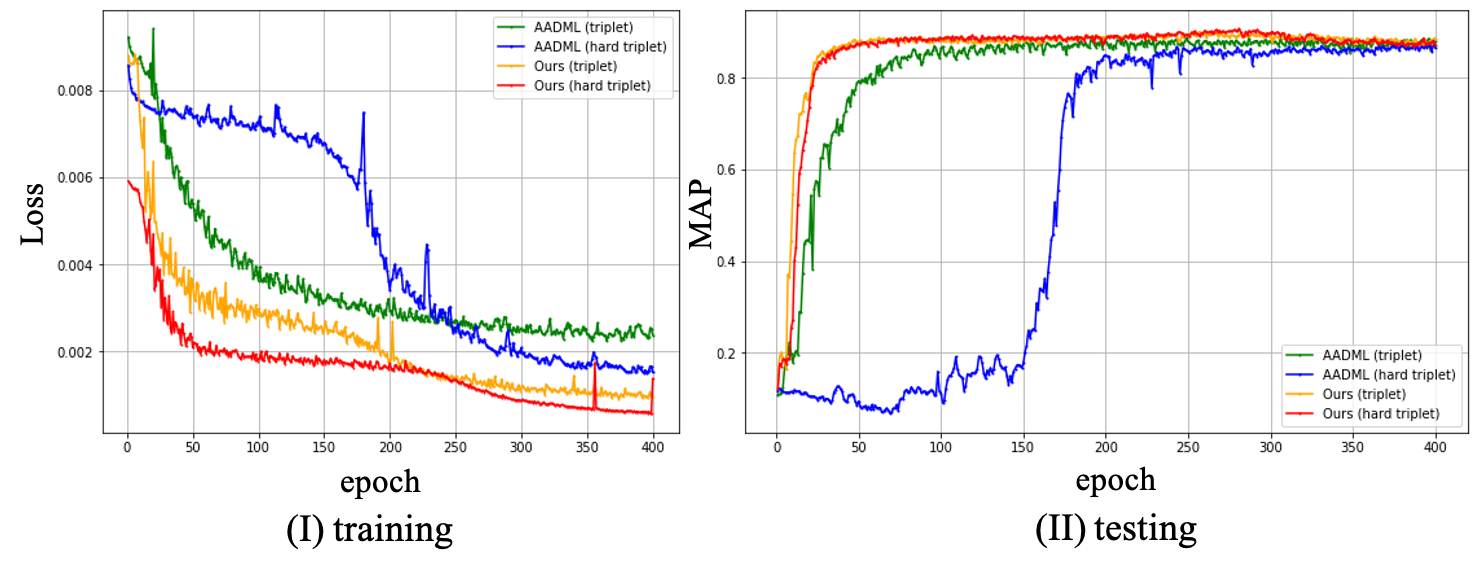}
    \caption{Comparison of loss and MAP over epochs on the AVE dataset between ours and AADML in triplet and hard triplet.}
    \label{fig:epochmap}
    \vspace{-0.8pc}
\end{figure}
\begin{table}[t]
\centering
\caption{Comparison of MAP of our method and its variations on the AVE dataset, with the highest scores in bold. 
% $Loss$ denotes our full loss function (Eq.\ref{eq:loss}).
}
\begin{NiceTabular}{l|c|c|c}
\hline
\textbf{Model} & A$\rightarrow$V & V$\rightarrow$A  & Average \\ \hline
\textbf{Ours ($Loss$ cf. Eq.\ref{eq:loss}))}          & \textbf{0.909} &\textbf{0.907} &\textbf{0.908} \\
$Loss$ without $l_{dis}$ &0.899 &0.893 &0.896 \\
$Loss$ without self-dis &0.884 &0.885 &0.884 \\ \hline
$Loss$ without $AA$ & 0.881 &0.880 & 0.881\\ 
$Loss$ without $l_{dis}$ and $AA$ & 0.856 &0.867 &0.862 \\ \hline
$Loss$ (linear) & 0.896 & 0.902 & 0.899 \\ 
$Loss$ (cosine-annealing) & 0.901 &0.903 &0.902 \\ \hline
\end{NiceTabular}
\label{tab:component}
\vspace{-1.5pc}
\end{table}
\subsection{Ablation Studies}
\subsubsection{Triplet Selection Strategy}
The triplet selection strategy significantly affects model performance by influencing the quality and diversity of training triplets. We compare the $\textit{triplet}$ strategies, which use all triplets within a batch, with the $\textit{hard triplet}$ strategy, which selects only the hardest triplet for model training. We evaluate these strategies with progressive self-distillation and compare their effectiveness to AADML, as shown in Fig.~\ref{fig:epochmap}. The results demonstrate that progressive self-distillation improves the performance of triplet-based methods over AADML in both training and testing. 
% , we compare training loss and testing MAP changes over epochs on the AVE dataset. The strategy $\textit{triplet}$ selects all possible triplets within a batch to train the model. Another strategy $\textit{hard triplet}$ only uses the hardest triplet consisting of the hardest negative for each anchor-positive pair. We evaluate these strategies in conjunction with progressive self-distillation and compare their effectiveness with AADML, as shown in Fig.~\ref{fig:epochmap}. Our experiments show that progressive self-distillation enhances triplet-based methods' performance compared to AADML, which outperforms it in both training and testing

% \begin{table}[ht]
% \centering
% \resizebox{\columnwidth}{!}{
% \begin{tabular}{|l|c|c|c|}
% \hline
% \textbf{Model} & \textbf{Audio Retrieval} & \textbf{Visual Retrieval} & \textbf{Average} \\
% \hline
% Triplet (w/) & 89.63 & 90.47 & 90.05 \\
% Triplet (w/o) & 88.37 & 89.49 &  88.93 \\
% \hline
% % Hard Triplet (w/) & \textbf{90.47} (1.46$\uparrow$) & 90.15 & \textbf{90.31} (1.64$\uparrow$) \\
% Hard Triplet (w/) & \textbf{90.87} (1.86$\uparrow$) & \textbf{90.71} (2.38$\uparrow$) & \textbf{90.79} (2.12$\uparrow$) \\
% Hard Triplet (w/o) & 89.71  & 89.18 & 89.45 \\ 
% \hline
% SOTA & 89.01 & 88.33 & 88.67 \\
% \hline
% \end{tabular}}
% \caption{MAP (\%) of Audio-Visual Retrieval with the change of factor 1}
% \label{table:discrepancy}
% \end{table}
\subsubsection{Impact of Different Components}
To elucidate the process for identifying the best model, we analyze several key factors and conduct a performance comparison. These factors include eliminating cross-modal dependency during training, the usage of $AA$ proxy~\cite{zeng2024anchor} for the sample representation, and the strategies for decreasing the $r$ gradually. As shown in Table~\ref{tab:component}, the best model incorporates cross-modal dependency elimination, the $AA$ proxy, and a step-wise strategy for decreasing $r$. We observe that cross-modal dependency elimination ($l_{dis}$) enhances performance with progressive self-distillation, but has a lesser effect without it. Compared to Table~\ref{tab:comparison}, the $AA$ proxy significantly improves the performance in the AADML model, but our method achieves nearly the best performance even without it.
% our method, even without the $AA$ proxy, achieves the second-best performance.
Finally, the step-wise strategy yields the best performance among the other strategies.

\section{Conclusion}
\label{sec:majhead}
Our approach overcomes the limitations of label-guided representation learning in existing metric learning methods by integrating cross-modal triplet loss with progressive self-distillation, enabling more robust audio-visual embedding learning. Our experiments demonstrate substantial improvements over existing methods in cross-modal retrieval tasks. Future work will extend this approach to other multimodal scenarios and explore real-world applications.

\bibliography{refs}
\bibliographystyle{plain}

\end{document}